# Secure Expansion of Energy Storage and Transmission Lines Considering Bundling Option Under Renewable Penetration

Mojtaba Moradi-Sepahvand, and Turaj Amraee, *Senior Member, IEEE*

*Abstract*—This paper presents a multi-stage expansion model for the co-planning of transmission lines, battery energy storage (ES), and wind power plants (WPP). High penetration of renewable energy sources (RES) is integrated into the proposed model concerning renewable portfolio standard (RPS) policy goals. The possibility of bundling existing transmission lines to uprate power flow capacity is considered. Renewable energy curtailment and load shedding are included in the model to assess the system operation more precisely. Battery ES devices are co-planned to defer transmission expansion and renewable management. To make the time complexity of the problem tractable and capture the uncertainties of load and RES in an hourly resolution, a chronological time-period clustering algorithm is used to extract the representative hours of each planning stage. Additionally, the flexible ramp reserve is utilized to handle the uncertainty of RES. An accelerated Benders dual decomposition (BDD) algorithm is developed to solve the proposed model mixed-integer linear programming (MILP) formulation. The *N-1* security criterion is evaluated by considering a designed contingency screening (CS) algorithm to identify higher risk contingencies. The effectiveness of the proposed co-planning model is evaluated using IEEE RTS 24-bus and IEEE 118-bus test systems.

## NOMENCLATURE

*Acronyms:*

| | |
|---|---|
| BDD | Benders dual decomposition. |
| CS, CTPC | Contingency screening, chronological time-period clustering. |
| CRF, EAC | Capital recovery factor, equivalent annual cost. |
| DPV, ES | Discounted present values, energy storage. |
| LB, UB | Lower bound, upper bound. |
| LI, NS | Loading index, *N-1* scenarios. |
| MILP, OPF, POC | Mixed-integer linear programming, optimal power flow, Pareto optimality cut. |
| MP, DSP | Master problem, dual sub-problem. |
| MDSP, CSDSP, NDSP | Modified dual sub-problem, contingency screening dual sub-problem, new dual sub-problem. |
| RES, RoW, RPS, WPP | Renewable energy sources, right of way, renewable portfolio standard, wind power plants. |
| TEP, TPC, TIC, TOC | Transmission expansion planning, total planning cost, total investment cost, total operation cost. |

*Indices & Sets:*

| | |
|---|---|
| $t, \Omega_T, T$ | Index, Set, and total number of planning stages. |
| $h, \Omega_H, H$ | Index, Set, and total number of representative hours. |
| $i, \Omega_B, \Omega_G$ | Index and Set of all buses, and set of generators. |
| $n, \Omega_N$ | Index and Set of the required number of conductors (two/four) per phase for bundling candidate existing lines. |
| $\Omega_{sb}, \Omega_w$ | Set of candidate buses for installing battery ES devices and constructing WPPs. |
| $l, \ell, L$ | Index of all lines, Index of single and double circuit candidate lines for bundling, and total number of lines. |
| $\Omega_{nl}, \Omega_{nc}$ | Sets of all candidate lines, and candidate lines in new corridors as a subset of $\Omega_{nl}$. |
| $\Omega_{el}, \Omega_{bl}$ | Sets of existing lines, and candidate lines for bundling. |
| $c, \Omega_C$ | Index and Set of the allowable candidates in a corridor. |
| $p, \Omega_P, P$ | Index, Set, and total number of linear segments of the generation cost function. |

*Parameters:*

| | |
|---|---|
| $r, LT$ | Interest rate and Lifetime of equipment (year). |
| $IC_l$ | Investment cost of new line *l* including the cost of conductors of single/double circuits and towers (M$/km). |
| $ICb_\ell$ | Investment cost of new bundled line $\ell$ including the cost of two/four conductors per phase for single/double circuit candidate lines and towers modification (M$/Km). |
| $ICw_i$ | Investment cost of new WPP (M$/MW). |
| $Rw_l$ | Right of Way cost for line *l* including land cost (M$/Km). |
| $LL_l, AS_l$ | Line length of all new lines (Km), and new substation cost in new corridors (M$). |
| $Cg_i^p$ | Cost of power generation in each segment *p* for each thermal unit *i* ($/MWh). |
| $Cls_i, Cwc_i$ | Load shedding and wind curtailment penalty cost ($/MWh) in bus *i*. |
| $RPU_i, RPD_i$ | Ramp up and ramp down limits of thermal unit *i* (MW). |
| $\alpha, \beta$ | Expected proportion of WPP in supplying the total load at the end of planning horizon, the maximum annual wind curtailment in the system. |
| $\gamma, \Phi$ | Maximum allowable hourly load shedding in each bus, and annual load shedding in the system as percentages of the bus load, and the total load. |
| $\vartheta$ | Energy to power ratio of battery ES (*hour*). |
| $\{\bullet\}^{max}, \{\bullet\}^{min}$ | Maximum/minimum limits of bounded variables. |
| $Ab, Ac$ | Discriminant matrices of selected existing lines for bundling, and percentage of power flow capacity increasing for bundled line due to two/four conductors per phase. |
| $A, K$ | Directional Connectivity matrices of existing and new lines with buses. |
| $Cs_i, Cc_i, dc_i$ | Investment cost of energy capacity ($/MWh), power |

M. Moradi-Sepahvand is with the Department of Electrical Sustainable Energy, TU Delft, Delft, The Netherlands (e-mail: m.moradisepahvand@tudelft.nl).
T. Amraee is with the Faculty of Electrical Engineering, K.N. Toosi University of Technology, Tehran, Iran (e-mail: amraee@kntu.ac.ir).

| | |
|---|---|
| | capacity ($/MW), and degradation cost ($/MWh) for each battery ES device in bus $i$. |
| $\eta_c, \eta_d$ | Charging and discharging efficiency of battery ES devices. |
| $Wf_h, Lf_h$ | Hourly representative factors obtained for wind power and load demand. |
| $Ld_i^{PK}, Lg$ | Peak load of bus $i$ (MW), Load Growth factor. |
| $\rho_h$ | The weight of obtained representative hour $h$. |
| $\chi$ | The reserve cost factor as a percentage of the cost of power generation. |
| $M, \Psi, B$ | Big-M, the base power of the system (MVA), per unit susceptance of all lines. |

*Variables:*

| | |
|---|---|
| $Z$ | Total Planning Cost. |
| $Y_{t,l,c}$ | Binary variables of candidate line $l$ at stage $t$ and corridor $c$ (equals 1 if the candidate line is constructed and 0 otherwise). |
| $Yb_{t,l}$ | Binary variables for bundling existing single/double circuit line $l$ at stage $t$. |
| $Pe_{t,l,h}$ | Flow of existing line $l$ at stage $t$ and hour $h$ (MW). |
| $I_{t,i,h}, U_{t,i,h}$ | On/off state of thermal unit $i$ at stage $t$, and hour $h$, Charging/discharging state of battery ES in bus $i$ at stage $t$, and hour $h$. |
| $S_{t,i}, C_{t,i}$ | Total energy (MWh) and power (MW) capacity of battery ES $i$ at stage $t$. |
| $P_{t,i,h}$, $Ps_{t,i,h,p}$ | Power output of thermal unit $i$ at stage $t$ and hour $h$, Power generation of segment $p$ of unit $i$ at stage $t$ and hour $h$ (MW). |
| $Pw_{t,i}, PC_{t,i,h}$ | Total power capacity of WPP $i$ at stage $t$ (MW), wind curtailment of WPP $i$ at stage $t$ and hour $h$ (MW). |
| $R_{t,i,h}, LS_{t,i,h}$ | Reserve of thermal unit $i$ at stage $t$ and hour $h$ (MW), load shedding in bus $i$ at stage $t$ and hour $h$ (MW) |
| $E_{t,i,h}$ | Stored energy (MWh) of battery ES in bus $i$ at stage $t$ and hour $h$. |
| $Pd_{t,i,h}, Pc_{t,i,h}$ | Discharging and charging power of battery ES in bus $i$ at stage $t$ and hour $h$ (MW). |
| $Pl_{t,l,c,h}$ | Power flow of new constructed line $l$, in corridor $c$, at stage $t$ and hour $h$ (MW). |
| $\theta_{t,i,h}$ | Voltage angle of bus $i$ at stage $t$, and hour $h$. |

*Compact Representation:*

| | |
|---|---|
| **Y** | Vector of binary decision variables. |
| **S** | Vector of battery ES power and energy capacity variables. |
| **W** | Vector of WPP power capacity variables. |
| **P** | Vector of positive continuous operational variables. |
| **Q** | Vector of free continuous variables. |

I. INTRODUCTION

*A. Background and Literature Review*

THE modern power systems are highly affected by the large integration of renewable energy sources (RES). Such a high penetration is caused due to the need for clean energy under different facilitating policies, such as renewable portfolio standard (RPS) policy, feed-in tariffs, and tradable green certificates. RPS is a successful policy for energy transition from conventional fossil-fuel generation units to RESs, such as wind and solar. Based on RPS policy, it is mandatory for energy utilities to get a predetermined amount of required energy from RESs, by a certain year. In this regard, the transmission expansion planning (TEP) that is essential for enhancing and improving the power transfer between remote RES and demand centers is affected [1]. The uncertainties of RES, like wind power plants (WPP), impact the TEP problem, and on the generation side, due to the RPS policy, the future generation mix is expected to be changed [2]. Most of the literature, e.g., [3-5], typically considers the WPPs as certain installed capacities that are available to be connected to some buses. Integrating the WPP optimal location and size is essential to conduct a power system planning study [6, 7]. Moreover, due to encouragement from governments on investment of renewables, the power system planning should be investigated under high-level of renewable penetration [8]. In this regard a power system expansion planning considering transmission lines and generation units is proposed in [8] to investigate the issues of a fully renewable generation mix in a long-term planning model.

Generally, uncertainties in the power system can be categorized as the uncertainty of technical and economic parameters. The technical uncertainty itself is divided into topological and operational uncertainties. The topological uncertainty indicates outage of elements (e.g., generator or transmission line) in the power system. The operational uncertainty can be expressed as load demand, load growth, and renewables output uncertainties. The economic uncertainty mainly indicates electricity market price variation, gross domestic product, and economic growth. In TEP problems, technical uncertainties are usually considered [2, 5, 6, 9-12]. To manage the uncertainties of WPP, mitigate transmission congestion, and defer the transmission expansion, ES devices can be utilized. The effect of ES devices on congestion caused by renewable integration has been investigated in [13]. In [14], a static security-constrained co-planning of the transmission and ES devices under high penetration of WPP is developed. In [4], a continuous time model for sitting and sizing of fast acting ES devices in transmission systems under penetration of WPP is presented to supply the fast ramping requirements. The co-planning of the transmission grid along with the location and capacity of ES, is addressed in [15] for reaching demand shifting and transmission upgrade deferrals. In [5], a stochastic dynamic co-planning of ES and transmission network is presented considering WPP and load uncertainty. In [16], a stochastic MILP model is proposed to assess the impacts of battery ES devices on the operation of transmission system under RES penetration. To obtain the optimal operation of ES devices for minimizing the operational cost of power system and maximizing the ES owner benefit, a bi-level strategy is introduced in [17]. In [18] a coordinated expansion planning model is proposed for both transmission and distribution systems in which ES and other smart grid technologies like electric vehicle taxis and demand response programs are considered. A bi-level optimization problem for combining operation of transmission system and energy management of independent substations in distribution system considering ES devices is introduced in [19].

In [7], WPP planning is added to the static TEP problem. Also, a unit commitment model is used to capture the operational aspects of load and generation changes. In [6], a coordinated static model is proposed for co-planning WPP, ES, and transmission network, with transmission switching to obtain a flexible network topology. In [9], both transmission and ES are planned to minimize the investment cost under the high penetration of WPP. The long-term uncertainty of future generation capacity and peak load in the target year, along with the short-term uncertainty of daily operating conditions are considered in [9]. An expansion planning model for power system is developed in [20] to handle high penetration of renewables considering the influence of ES devices. A bi-level formulation is developed in [21] to model a multi-energy planning model considering energy markets and the sitting and sizing of electrical battery ES devices.

Although ES devices have an impressive impact on avoiding new transmission lines construction, the high cost of new right of way (RoW) and environmental issues encourage the planners to use the existing towers and RoW to uprate the capacity of transmission lines. Transmission line bundling is one of the options to use the existing structure for increasing the power flow of transmission lines and decreasing the power losses [10]. By bundling the transmission lines, two or more parallel conductors are used instead of a single conductor. Therefore, the voltage gradient at the surface of conductors is minimized. In the high voltage range, two, three, or four conductors per phase are commonly used for bundling, while the ultra-high voltage range may have six, eight, or twelve conductors. The most significant advantages of transmission line bundling are reducing lines reactance and resistance, along with corona loss, and enhancing lines power flow by increasing the current carrying capacity [22]. In [23], a bundling geometry is proposed to optimize the transmission line capacity by which geometries with a higher surge impedance loading than the original, under a reduced cost and a smaller RoW are obtained in [23]. In [24], a phase sequence optimization method is presented to reach multi-circuit transmission lines with increased SIL, and more limited RoW, but with the same cost as the conventional lines. A mathematical model for bundle geometry and conductor type optimization is proposed and evaluated in [25] to enhance the SIL, and decrease the cost, RoW, and height of transmission line towers. In [10], the number of bundled conductors for candidate lines is optimized to evaluate the short circuit level in a short-circuit constrained system expansion planning model. Therefore, upgrading the existing lines using bundling is a potential option in the TEP study.

A major challenge in co-planning transmission, WPP, and ES devices, is the accuracy of the system operation modeling. To capture the uncertainty of load demand, wind power generation, and daily cycles of ES in an hourly time resolution, efficient methods should be developed for extracting representative intervals in each planning year [11]. In most of the recent investigations, a stochastic programming framework is considered to represent the uncertainties. In stochastic programming, a large number of scenarios are used to represent the uncertainties accurately. Considering a large number of scenarios increases computational complexity. In this regard, extracting effective representations is essential. For extracting representations, the common techniques are K-means ([5], [6], and [9]) and hierarchical clustering ([11, 26]). The accuracy of previous methods in capturing uncertainty

TABLE I
AN OVERVIEW OF THE PREVIOUS WORKS

| Ref[1] | Overall Model Description | | | | | | |
|---|---|---|---|---|---|---|---|
| | DPH[2] | ESD[3] | WPP[4] | SUC[5] | TLB[6] | NSC[7] | CSA[8] |
| [3] | ✓ | ✓ | --- | ✓ | --- | --- | --- |
| [4], [9], [13] | --- | ✓ | --- | ✓ | --- | --- | --- |
| [5] | ✓ | ✓ | --- | ✓ | --- | --- | --- |
| [6] | --- | ✓ | ✓ | ✓ | --- | --- | --- |
| [7] | --- | ✓ | ✓ | --- | --- | --- | --- |
| [8] | ✓ | --- | ✓ | ✓ | --- | ✓ | --- |
| [10] | ✓ | --- | --- | --- | ✓ | --- | --- |
| [11] | ✓ | --- | --- | ✓ | --- | --- | --- |
| [14] | --- | ✓ | --- | ✓ | --- | ✓ | --- |
| [20] | ✓ | ✓ | ✓ | ✓ | --- | --- | --- |
| [21] | --- | ✓ | --- | ✓ | --- | --- | --- |
| [27], [28], [29] | --- | --- | ✓ | ✓ | --- | --- | --- |
| Proposed Model | ✓ | ✓ | ✓ | ✓ | ✓ | ✓ | ✓ |

**1**: References, **2**: Dynamic Planning Horizon, **3**: Energy Storage Devices, **4**: Wind Power Plant, **5**: Short-term Uncertainty Capturing, **6**: Transmission Line Bundling, **7**: N-1 Security Criterion, **8**: Contingency Screening Algorithm.

may decrease under the high integration of RES and ES [12]. Moreover, in some studies scenario generation approaches are utilized to incorporate important scenarios in TEP problem [27, 28]. In [27] a scenario generation method is presented to model load and WPP scenarios in TEP problem considering inter-spatial correlation between data. To consider many scenarios in TEP problem, a method is suggested in [28], which is based on cost-oriented dynamic scenario clustering. In [28], the operation sub-problems of utilized Benders decomposition (BD) are clustered to reduce the complexity. In order to have a tradeoff between cost, robustness, and computational burden of TEP problem, an approach considering uncertainty budget as a variable for WPPs output intermittency is investigated in [29]. In [27], [28], and [29] no ES device is incorporated in TEP problem, and their presented models are static ignoring a more realistic dynamic planning horizon.

### B. Research Gaps and Contributions

To discuss the research gaps, the previously reviewed papers in background and literature review subsection are summarized and listed in Table I. The majority of previous research relies on installing new transmission lines for integrating certain installed capacities of RES ignoring ES devices. The planning is also conducted for a given target year without considering the installation timing of new devices. Another gap in previous research is the lack of an efficient clustering approach for extracting representative hours in each planning stage to capture the short-term operational aspects of generating units, load demand, WPPs, and cycling of ES devices. There are major differences between this work and the previous TEP problems. In this paper, a multi-stage secure model for the co-planning of transmission lines, battery ES devices, and WPPs is proposed concerning RPS policy goals. Unlike previous studies, both battery ES devices and upgrading existing transmission lines using the bundling option are utilized to maximize the integration of RES in the transmission network. This is the first time that a secure co-planning model is presented to plan the battery ES devices and bundle the existing lines to promote the transmission expansion and simultaneously the renewable integration. Moreover, renewable energy curtailment and load shedding are also included to model the system operation accurately. There are two terms of reliability, including adequacy and security. The focus of this paper is on the security criterion. While the proposed model is multi-stage and operational details are incorporated, the *N-1* security criterion is considered using a heuristic contingency screening (CS) algorithm to identify the higher risk contingencies. Indeed, both battery ES devices and bundling help to reach a secure plan, and this issue has not been addressed in previous studies. In addition, to precisely capture short-term uncertainties of load demand and renewable output power in a long-term multistage planning model, an accurate chronological time-period clustering (CTPC) algorithm, along with some metrics for finding the suitable number of representative hours is utilized. Finally, such a comprehensive model needs a sophisticated solution algorithm to reach the global optimal solution. A customized and accelerated Benders dual decomposition (BDD) algorithm is developed for the proposed model, while in previous studies, this solution algorithm has not been presented. Note that in this paper no expansion planning in the generation side is considered. In what follows, the main contributions of this work are summarized.

1) A multi-stage secure model for the transmission line, battery ES device, and WPP co-planning is presented concerning RPS policy goals. The battery ES devices contribute to uncertainty handling of renewable and load demand as well as relieving transmission congestion and expansion deferral. The possibility of bundling existing transmission lines to increase power flow capacity is considered. Renewable energy curtailment and load shedding are also included in the model for accurate modeling of system operation. The *N-1* security criterion is evaluated using a heuristic CS algorithm to identify the higher risk contingencies.

2) A methodology based on the accelerated BDD algorithm for dealing with the complexity of the MILP formulation of the proposed multi-stage co-planning problem is developed. The proposed method handles the *N-1* contingency analysis in a decomposed structure. In addition, a CTPC algorithm is used to extract the representative hours to manage the time complexity of the proposed co-planning model and accurately capture the uncertainties of load and RES.

The rest of the paper is organized as follows. In Section II, the detailed formulation of the proposed model is given. The proposed model overall structure and the CTPC algorithm for extracting the representative hours are presented in Sections III and IV, respectively. The simulation results of the proposed method over the test system are discussed in Section V. Finally, the paper is concluded in Section VI.

## II. FORMULATIONS

The formulations of the proposed co-planning model according to DC optimal power flow (OPF) are presented in general and BDD forms. The objective function and the related constraints in the general form are described as follows.

### A. Objective Function

Based on (1), the objective function minimizes the discounted present values (DPV) of the total investment cost (TIC), and the total operation cost (TOC). It should be noted that for the sake of simplicity, maintenance and decommissioning of equipment are not modeled. The DPV of the TIC is considered at the beginning of each planning stage, which is assumed 2-years, and for the TOC, it is considered at the end of each planning stage. The (1a) refers to the TIC and is equal to the DPV cost of all new constructed transmission lines, bundled lines, battery ES devices, and WPP. The (1b) represents the TOC. It includes the linearized cost function of thermal generating units with required flexible ramp reserve cost, the degradation cost of battery ES, and the total load shedding and wind curtailment costs. The annual DPV investment cost is calculated using capital recovery factor (CRF), which converts the DPV to equivalent annual cost (EAC) [30].

$$Min \ \ Z = TIC + TOC \tag{1}$$

$$TIC = \sum_{t \in \Omega_T} \ \ [\left(\frac{2}{(1+r)^{2t-1}}\right) \times ( \tag{1a}$$

$$\frac{10^6 \times r(1+r)^{LT_{Line}}}{(1+r)^{LT_{Line}}-1} \times \left[\sum_{l \in \Omega_{nl}} IC_l \ . LL_l \times (\sum_{c \in \Omega_C} Y_{t,l,c}) + \right.$$

$$\sum_{l \in \Omega_{nl}} [Rw_l . LL_l \times \sum_{c \in \Omega_C} Y_{t,l,c}] + \sum_{l \in \Omega_{nc}} [Y_{t,l,c=1} . (AS_l)] ] +$$

$$\frac{10^6 \times r(1+r)^{LT_{Line}}}{(1+r)^{LT_{Line}}-1} \times LL_\ell \times \left[\sum_{\ell \in \Omega_{bl}} ICb_\ell \ . Yb_{t,\ell}\right] +$$

$$\sum_{i \in \Omega_{sb}} \left[Cs_i \ \frac{r(1+r)^{LT_{ES}}}{(1+r)^{LT_{ES}}-1} \times (S_{t,i}) + Cc_i \ \frac{r(1+r)^{LT_{ES}}}{(1+r)^{LT_{ES}}-1} \times (C_{t,i})\right] +$$

$$\frac{10^6 \times r(1+r)^{LT_{WPP}}}{(1+r)^{LT_{WPP}}-1} \times \left[\sum_{i \in \Omega_W} ICw_i \ . Pw_{t,i}\right] \ ) ]$$

$$TOC = \sum_{t \in \Omega_T} \ \ [\left(\frac{2}{(1+r)^{2t}}\right) \times [8760 \times \sum_{h \in \Omega_H} \rho_h \times ( \tag{1b}$$

$$\sum_{i \in \Omega_G} [[\ Cg_i^{p=1} \ . (P_i^{min} . I_{t,i,h} + \chi . R_{t,i,h})] + \sum_{p \in \Omega_P} [Cg_i^p . Ps_{t,i,h,p}]] +$$

$$\sum_{i \in \Omega_{sb}} dc_i \ . Pd_{t,i,h} +$$

$$\sum_{i \in \Omega_B} Cls_i \ . LS_{t,i,h} + \sum_{i \in \Omega_w} Cwc_i \ . PC_{t,i,h} \ ) ]$$

The EAC of new constructed lines in existing or new corridors is formulated as the first term of TIC. In all corridors, the RoW cost is considered for single and double circuit lines, and in new corridors, the substation cost is considered only for the first new corridor. The EAC cost for bundling the existing lines, including the cost of two or four bundled conductors for single/double circuit existing lines and tower upgrading, is formulated as the second term of TIC. The EAC investment cost for battery ES devices and WPPs is represented by the third and fourth terms of TIC, respectively.

The linearized cost function of thermal units and the required flexible ramp reserve cost is defined as the first term of TOC. The battery ES degradation cost, caused by the charging/discharging cycles and aging [5], is formulated as the second term of TOC based on the discharging power in each cycle. In this regard, a fixed degradation cost is assumed to simplify the battery ES degradation model. Finally, the total load shedding and wind curtailment penalty cost is presented in the last term of TOC. It should be noted that when wind energy is curtailed, the generation of thermal resources is increased to supply the load demand. In order to avoid wind curtailment as much as possible, a separate cost is included as wind curtailment penalty cost in TOC. Moreover, ES devices and constructing new transmission lines facilitate the WPP integration and minimize the wind curtailment. In other words, considering a wind curtailment penalty cost in the objective function motivates the utilization of ES devices to reduce renewable curtailment, and transmission congestion, and can defer the transmission investment.

### B. Technical and Economic Constraints

To model the total generation cost and deal with the technical and economic limits of generators, the constraints in (2) to (5) are introduced. The bounds of thermal units output power are considered using (2). In (3), the nonlinear cost function of thermal units is linearized. In this constraint, the hourly power generation of each thermal unit is assumed as the summation of the minimum power and a set of linear segments of generated power. In (4), the limits of power generation for each linearized segment are expressed. The ramping constraints of thermal units are introduced in (5).

$$P_i^{min}.I_{t,i,h} \leq P_{t,i,h} \leq P_i^{max}.I_{t,i,h} \quad \forall i \in \Omega_G, t \in \Omega_T, h \in \Omega_H \quad (2)$$

$$P_{t,i,h} = P_i^{min}.I_{t,i,h} + \sum_{p=1}^{P} Ps_{t,i,h,p} \quad \forall i \in \Omega_G, t \in \Omega_T, h \in \Omega_H \quad (3)$$

$$0 \leq Ps_{t,i,h,p} \leq (P_i^{max} - P_i^{min}).\frac{I_{t,i,h}}{P} \quad (4)$$
$$\forall i \in \Omega_G, t \in \Omega_T, h \in \Omega_H, p \in \Omega_P$$

$$\begin{cases} P_{t,i,h} - P_{t,i,h-1} \leq RPU_i & \forall i \in \Omega_G, t \in \Omega_T, h \in \Omega_H \\ P_{t,i,h-1} - P_{t,i,h} \leq RPD_i & \forall i \in \Omega_G, t \in \Omega_T, h \in \Omega_H \end{cases} \quad (5)$$

To model the RPS policy, the constraints in (6) to (8) are introduced. In (6), the upper and lower bounds of the WPP power capacity are defined. The minimum installed WPP to supply the load is assumed as a percentage of the total peak load at each planning stage. According to the RPS policy, it is assumed that at the last stage of the planning horizon, the total capacity of installed WPP will be 15% of the total peak load, as expressed in (7). Based on (8), the installed WPP in a given stage will remain in the system till the end of the planning horizon.

$$0 \leq Pw_{t,i} \leq Pw_i^{max} \quad \forall i \in \Omega_w, t \in \Omega_T \quad (6)$$

$$[\alpha \times t/T] \times (1+Lg)^{2t}.\sum_{i \in \Omega_B} Ld_i^{pk} \leq \sum_{i \in \Omega_w} Pw_{t,i} \quad \forall t \in \Omega_T \quad (7)$$

$$Pw_{t-1,i} \leq Pw_{t,i} \quad \forall i \in \Omega_w, t \in \Omega_T \quad (8)$$

A variety of reasons such as the intermittency of WPP, transmission line congestion, insufficient transmission access, and surplus power generation during times of low load demand, can lead to wind curtailment. Therefore, modeling the wind curtailment and the load shedding is an essential part of the TEP problem, especially in the presence of ES devices. In (9), the limits of hourly wind curtailment in each WPP bus are specified. Based on (10), the maximum annual wind curtailment in the system is assumed as a certain percentage of the expected power output of the WPP. The limits of hourly allowable load shedding in each bus are defined using (11). The upper limit of load shedding is expressed as a certain percentage of the expected load. The maximum annual allowable load shedding in the system is considered as a certain percentage of the total expected load, as given in (12).

$$0 \leq PC_{t,i,h} \leq Wf_h.Pw_{t,i} \quad \forall i \in \Omega_w, t \in \Omega_T, h \in \Omega_H \quad (9)$$

$$\sum_{i \in \Omega_w} \sum_{h \in \Omega_H} PC_{t,i,h} \leq \beta \times \sum_{i \in \Omega_w} \sum_{h \in \Omega_H} Wf_h.Pw_{t,i} \quad \forall t \in \Omega_T \quad (10)$$

$$0 \leq LS_{t,i,h} \leq \gamma.(1+Lg)^{2t}.Lf_h.Ld_i^{pk} \quad \forall i \in \Omega_B, t \in \Omega_T, h \in \Omega_H \quad (11)$$

$$\sum_{i \in \Omega_B} \sum_{h \in \Omega_H} LS_{t,i,h} \leq \Phi \times (1+Lg)^{2t}.\sum_{i \in \Omega_B} \sum_{h \in \Omega_H} Lf_h.Ld_i^{pk} \quad (12)$$
$$\forall t \in \Omega_T$$

In order to handle the uncertainty of load and WPP, the flexible ramp reserve is modeled as given in (13) to (15). It should be noted that in this paper, both downward and upward flexible ramp reserves are assumed to be the same. For each thermal unit, the reserve limits are represented by (13). The summation of reserve and the output power of each unit is restricted by (14). In (15), the minimum limit for the total hourly flexible ramp reserve is considered as 5% of the total expected WPP output power plus 3% of the system load [3].

$$0 \leq R_{t,i,h} \leq P_{t,i,h} \quad \forall i \in \Omega_G, t \in \Omega_T, h \in \Omega_H \quad (13)$$

$$R_{t,i,h} + P_{t,i,h} \leq P_i^{max} \quad \forall i \in \Omega_G, t \in \Omega_T, h \in \Omega_H \quad (14)$$

$$\sum_{i \in \Omega_G} R_{t,i,h} \geq (5\%) \times \sum_{i \in \Omega_w} Wf_h.Pw_{t,i} \quad (15)$$
$$+(3\%) \times (1+Lg)^{2t}.Lf_h.\sum_{i \in \Omega_B} Ld_i^{pk} \quad \forall t \in \Omega_T, h \in \Omega_H$$

By integrating large-scale WPP, ES planning, i.e., sitting and sizing, is a key tool for reducing transmission congestion, wind curtailment, and load shedding. Constraints in (16) and (17) determine the limits of charging and discharging power of battery ES. In (18) and (19), the hourly status of battery ES charging and discharging is determined. In (20), at each hour of operation, the level of stored energy in battery ES devices is the stored energy at the previous hour plus the energy exchange at the current hour. The energy to power ratio of battery ES devices is represented using (21). The limits of the energy level with power and energy capacity of battery ES devices are considered in (22) to (24), respectively. The constraints (25) and (26) confirm that the installed battery ES in a given stage will remain in the system till the end of planning horizon.

$$0 \leq \eta_c.Pc_{t,i,h} \leq C_{t,i} \quad \forall i \in \Omega_{sb}, t \in \Omega_T, h \in \Omega_H \quad (16)$$

$$0 \leq 1/\eta_d.Pd_{t,i,h} \leq C_{t,i} \quad \forall i \in \Omega_{sb}, t \in \Omega_T, h \in \Omega_H \quad (17)$$

$$\eta_c.Pc_{t,i,h} \leq C_i^{max}.U_{t,i,h} \quad \forall i \in \Omega_{sb}, t \in \Omega_T, h \in \Omega_H \quad (18)$$

$$1/\eta_d . Pd_{t,i,h} \leq C_i^{max}.(1 - U_{t,i,h}) \quad \forall i \in \Omega_{sb}, t \in \Omega_T, h \in \Omega_H \quad (19)$$

$$E_{t,i,h} = E_{t,i,h-1} + \rho_h \times \left((\eta_c . Pc_{t,i,h}) - (1/\eta_d . Pd_{t,i,h})\right) \quad (20)$$
$$\forall i \in \Omega_{sb}, t \in \Omega_T, h \in \Omega_H$$

$$C_{t,i}.\vartheta \leq S_{t,i} \quad \forall i \in \Omega_{sb}, t \in \Omega_T \quad (21)$$

$$0 \leq E_{t,i,h} \leq S_{t,i} \quad \forall i \in \Omega_{sb}, t \in \Omega_T, h \in \Omega_H \quad (22)$$

$$0 \leq C_{t,i} \leq C_i^{max} \quad \forall i \in \Omega_{sb}, t \in \Omega_T \quad (23)$$

$$0 \leq S_{t,i} \leq S_i^{max} \quad \forall i \in \Omega_{sb}, t \in \Omega_T \quad (24)$$

$$S_{t-1,i} \leq S_{t,i} \quad \forall i \in \Omega_{sb}, t \in \Omega_T \quad (25)$$

$$C_{t-1,i} \leq C_{t,i} \quad \forall i \in \Omega_{sb}, t \in \Omega_T \quad (26)$$

To avoid constructing new transmission lines, the possibility of bundling existing single and double circuit lines with two or four conductors per phase is considered as a transmission uprate solution. The constraint given in (27) represents each existing line power flow considering the possibility of bundling. Accordingly, if an existing line is considered for bundling, the flow of that line is calculated using (28), which determines the percentage of line flow uprate. In (29), the limits of existing line flow regarding bundling possibility are represented. The constraint of (30) guarantees that the bundled lines in each stage will remain in the system until the end of the planning horizon.

$$-M_l . \sum_{\ell \in \Omega_{bl}} Ab_\ell^l . Yb_{t,\ell} \leq Pe_{t,l,h} - \sum_{i \in \Omega_B} \Psi . B_l . A_i^l . \theta_{t,i,h} \leq$$
$$M_l . \sum_{\ell \in \Omega_{bl}} Ab_\ell^l . Yb_{t,\ell} \quad \forall l \in \Omega_{el}, t \in \Omega_T, h \in \Omega_H \quad (27)$$

$$-M_l . (1 - Yb_{t,\ell}) \leq \sum_{l \in \Omega_{el}} Ab_\ell^l . Pe_{t,l,h} - \sum_{i \in \Omega_B} \sum_{n \in \Omega_N} [(1 + Ac_\ell^n) . \Psi . B_\ell . A_i^\ell . \theta_{t,i,h}] \leq M_l . (1 - Yb_{t,\ell}) \quad (28)$$
$$\forall \ell \in \Omega_{bl}, t \in \Omega_T, h \in \Omega_H$$

$$-P_l^{max} \times [1 + \sum_{\ell \in \Omega_{bl}} [(Ab_\ell^l . Yb_{t,\ell}) \times \sum_{n \in \Omega_N} Ac_\ell^n]] \leq Pe_{t,l,h} \leq$$
$$P_l^{max} \times [1 + \sum_{\ell \in \Omega_{bl}} [(Ab_\ell^l . Yb_{t,\ell}) \times \sum_{n \in \Omega_N} Ac_\ell^n]] \quad (29)$$
$$\forall l \in \Omega_{el}, t \in \Omega_T, h \in \Omega_H$$

$$Yb_{t-1,\ell} \leq Yb_{t,\ell} \quad \forall \ell \in \Omega_{bl}, t \in \Omega_T \quad (30)$$

The constraints in (31) and (32) refer to the power flow and the new candidate line limits, respectively. The first constraint in (33) guarantees that if a candidate line is constructed at a given stage, it is available at the next stages. The second constraint in (33) introduces the construction order of equivalent parallel lines.

$$-M_l . (1 - Y_{t,l,c}) \leq Pl_{t,l,c,h} - \sum_{i \in \Omega_B} \Psi . B_l . K_i^l . \theta_{t,i,h} \leq$$
$$M_l . (1 - Y_{t,l,c}) \quad \forall l \in \Omega_{nl}, t \in \Omega_T, c \in \Omega_C, h \in \Omega_H \quad (31)$$

$$-P_l^{max} . Y_{t,l,c} \leq Pl_{t,l,c,h} \leq P_l^{max} . Y_{t,l,c}$$
$$\forall l \in \Omega_{nl}, t \in \Omega_T, c \in \Omega_C, h \in \Omega_H \quad (32)$$

$$Y_{t-1,l,c} \leq Y_{t,l,c} \ \& \ Y_{t,l,c+1} \leq Y_{t,l,c} \quad \forall l \in \Omega_{nl}, c \in \Omega_C, t \in \Omega_T \quad (33)$$

In (34), the nodal power balance is defined. It includes the output power of thermal units and WPP considering wind curtailment, power exchange of ES devices, power flow of existing and new constructed lines, the total load, and load shedding in each bus.

$$P_{t,i,h} + [Wf_h . Pw_{t,i} - PC_{t,i,h}] + [Pd_{t,i,h} - Pc_{t,i,h}] -$$
$$\sum_{l \in \Omega_{el}} A_i^l . Pe_{t,l,h} - \sum_{l \in \Omega_{nl}} \sum_{c \in \Omega_C} K_i^l . Pl_{t,l,c,h} =$$
$$((1 + Lg)^{2t} . Lf_h . Ld_i^{PK}) - LS_{t,i,h} \quad \forall i \in \Omega_B, t \in \Omega_T, h \in \Omega_H \quad (34)$$

### C. Benders Dual Decomposition

In order to solve mixed-integer nonlinear programming (MINLP) optimization models, heuristic or evolutionary approaches are preferred since the mathematical algorithms (e.g., standard branch and bound) are not capable of finding the global optimal solution. However, when the optimization problems are expressed as LP or MILP models, powerful mathematical solution methods such as CPLEX algorithm can find the global optimal solution. In critical power system studies where the decision variables are very costly, the MILP models and related solvers are preferred. Decomposition algorithms are preferred in TEP studies where different expansion tools as binary variables with different cost terms are considered. Regarding these facts, and to model the *N-1* security criterion, in this paper a BDD-based solution methodology is developed for the proposed TEP model. Moreover, using the developed BDD algorithm makes it possible to assess different contingency scenarios during the planning horizon. The developed BDD algorithm is modeled in a duality environment due to its capability in making the searching space independent from integer variables and constructing more valid cuts.

In this part, the presented MILP formulations in (1) to (34) are reformulated using the BDD algorithm. The problem is decomposed into a master problem (MP) and a main dual sub-problem (DSP). It should be noted that DSP is indeed the dual form of sub-problem in which there is no integer decision variable. In this paper, just the formulation of DSP is presented directly to avoid extra formulation. In MP, the integer decision variables are optimized, and in DSP, the feasibility or optimality of MP solution and optimization of the system operation, WPP and ES investment costs are evaluated. In other words, MP obtained integer decision variables will be considered as constant parameters in DSP. In the following, an identical compact form for the objective function (1) and the constraints (2)-(34) is defined.

$$Min \ I_L^T Y + I_S^T S + I_W^T W + O_C^T P \quad (35)$$
s.t.

$$AY \geq B \quad (36)$$

$$CW + DP + EQ = F \quad : \sigma \quad (37)$$

$$G_1 Y + H_1 S + J_1 W + K_1 P + L_1 Q = M \quad : \lambda \quad (38)$$

$$G_2 Y + H_2 S + J_2 W + K_2 P + L_2 Q \geq N \quad : \mu \quad (39)$$

$$Y \in \{0,1\}, \quad S, W \ \& \ P \geq 0, \quad Q: free$$
$$Y = \{Y, Yb, U, I\}, \quad Q = \{\theta, Pl, Pe\}$$
$$P = \{P, Ps, R, Pd, Pc, E, PC, LS\}, \quad S = \{S, C\}, \quad W = \{Pw\}$$
$$\sigma \ \& \ \lambda: free, \quad \mu \geq 0$$

The objective function in (35) represents the objective function given by (1). The constraint in (36) denotes the constraints of (30) and (33). The equality constraint in (37) models (34). Constraint (38) corresponds to (3) and (20). The constraint given in (39) represents the constraints of (2), (4)-(19), (21)-(29), (31), and (32). The compact dual variables $\sigma, \lambda$ and $\mu$ are defined for the constraints (37), (38) and (39), respectively. $I_L$, $I_S$ and $I_W$ are the vectors for investment cost and $O_C$ is the vector of operation cost. The coefficients of $A, B, C, D, E, F, G_1, G_2, H_1, H_2, J_1, J_2, K_1, K_2, L_1, L_2, M$ and $N$ are all relevant matrices.

- *Master Problem*

The integer programming formulation of MP is expressed as follows:

$$Min \ Z_{lower} \quad (40)$$

s.t.

$$Z_{lower} \geq I_L^T Y \quad (41)$$

$$Z_{lower} \geq I_L^T Y + [F^T \bar{\sigma} + M^T \bar{\lambda} + N^T \bar{\mu}]^{(v)} + \bar{\pi}^{(v)}.(Y - \bar{Y}^{(v-1)}) \quad (42)$$

$$[M^T \bar{\lambda} + N^T \bar{\mu} + F^T \bar{\sigma}]^{(v)} + \bar{\pi}^{(v)}.(Y - \bar{Y}^{(v-1)}) \leq 0 \quad (43)$$

& (36)

The (40) is MP objective function which is the lower bound (LB) of the problem. The constraint of (41) represents the investment cost of binary decision variables, and the optimality and feasibility cuts are defined using (42) and (43). The iteration number is $v$, and $\pi$ is the dual variable of (44) as an auxiliary constraint for the sub-problem. In this constraint, the obtained binary decision variables from MP ($\bar{Y}$) are dedicated to a positive variable ($Y_{sp}$).

$$IY_{sp} = \bar{Y} \qquad : \pi$$
$$I: Identity \ Matrix, \quad Y_{sp} \geq 0, \quad \pi: free \quad (44)$$

- *Dual Sub-Problem*

The linear programming formulation of DSP is represented by (45) to (50).

$$Max \ F^T \sigma + M^T \lambda + N^T \mu + \bar{Y}^T \pi \quad (45)$$

s.t.

$$D^T \sigma + K_1^T \lambda + K_2^T \mu \leq O_C \quad \rightarrow \ : P \quad (46)$$

$$H_1^T \lambda + H_2^T \mu \leq I_S \quad \rightarrow \ : S \quad (47)$$

$$C^T \sigma + J_1^T \lambda + J_2^T \mu \leq I_W \quad \rightarrow \ : W \quad (48)$$

$$G_1^T \lambda + G_2^T \mu + I\pi \leq 0 \quad \rightarrow \ : Y \quad (49)$$

$$E^T \sigma + L_1^T \lambda + L_2^T \mu = 0 \quad \rightarrow \ : Q \quad (50)$$

The solution of MP determines the integer decision variables (i.e., $\bar{Y}$, that are considered as parameters in the DSP) to be used for obtaining the DSP solution. If the DSP solution is bounded, the optimality cut of (42) is constructed and the upper bound (UB) is calculated as follows.

$$UB = F^T \sigma + M^T \lambda + N^T \mu + \bar{Y}^T \pi + I_L^T \bar{Y} \quad (51)$$

Otherwise, if the solution is unbounded, the feasibility cut of (43) is generated using the following modified DSP (MDSP).

- *Modified DSP*

In order to deal with unbounded conditions in DSP and remove the extreme rays, an MDSP is introduced. Its objective function is assumed as (45), and its constraints are (46) to (50) all with right-hand-side equal to zero. Moreover, the (52), as an auxiliary constraint, is added.

$$\sigma \leq 1 \quad \& \quad \pi \leq 1 \quad (52)$$

At the end of each iteration, the algorithm is ended if the predefined tolerance in (53) is satisfied.

$$\frac{(UB-LB)}{UB} \leq \tau \quad (53)$$

- *Contingency Screening DSP*

The contingency screening DSP (CSDSP) is defined to evaluate the impact of each possible *N-1* contingency. The structure of CSDSP is similar to the DSP in which the load shedding is unlimited in each bus.

- *Acceleration Tools*

In order to accelerate the proposed BDD algorithm, the Pareto optimality cut (POC), inspired by the presented concepts in [31], and multiple cuts using the multiple-solution technique for MP, inspired by the presented outcomes in [32], are modeled. It should be noted that in both mentioned references the solution methodology is a classic Benders decomposition dealing with facility location problems, while in this paper a developed BDD algorithm is presented for solving a completely different problem. The POC is a strong and dominant optimality cut calculated after obtaining a bounded solution for the DSP. To obtain the POC, another DSP, i.e., new DSP (NDSP), is defined. In this NDSP, the objective function is as (45) except that $\bar{Y}^T$ is replaced by $\hat{Y}^{T(v)}$, which is the vector of core points for $Y$ in iteration $v$ as expressed in (54).

$$\hat{Y}^{T(v)} = \frac{1}{2}.\hat{Y}^{T(v-1)} + \frac{1}{2}.\bar{Y}^T \quad (54)$$

The constraints of the NDSP are like (46) to (50) plus the equality constraint in (55).

$$F^T \sigma + M^T \lambda + N^T \mu + \bar{Y}^T \pi = \mathbb{Z}_{DSP} \quad (55)$$

The right-hand-side of (55) is the obtained value for the objective function (45). After solving the NDSP, the optimality cut of (42) is constructed. POC can reduce the BDD algorithm iterations [31]. In addition, using the "solution pool" characteristic of the solver CPLEX in GAMS for MILP problems, after solving MP, multiple feasible solutions are obtained. Therefore, in each iteration, the DSP, MDSP, and NDSP for producing POC are solved for all obtained feasible solutions. Consequently, in each iteration, multiple cuts are generated. Having an MP with multiple cuts in each iteration, especially when the number of iterations is intensive, can lead to a significant improvement [32]. In multiple solution case, the LB is equal to the first solution of MP because it is the best solution, and UB is calculated based on a solution that has the minimum value for (45).

## III. OVERALL STRUCTURE OF THE PROPOSED MODEL

According to the accelerated BDD formulation, the overall structure of the proposed model is illustrated in Fig. 1. The model is decomposed to an MP, containing the binary variables, and four DSP (i.e., DSP, MDSP, CSDSP, and NDSP), including the continuous variables, to be solved using the BDD algorithm. According to Fig. 1, all the input data and initial values are defined, and then the BDD algorithm is started. In each iteration, the proposed CS algorithm is executed to determine the high-risk *N-1* scenarios (NSs), including outages of existing, new constructed, and bundled lines. It should be noted, in the first scenario, there is no contingency (i.e., normal or *N-0* scenario). The hourly power output of thermal units and WPP, the flexible ramp reserve requirement, the load shedding and wind curtailment, hourly power exchanges of ES, power flow of all lines, and the yearly capacity of ES and WPP are optimized in the DSP. In Fig. 1, 'a' is a flag variable with 0 or 1 values. The initial value for 'a' is equal to 0. After solving the DSP for each NS, if the solution is unbounded, 'a' is set to 1, and the MDSP (i.e., (45), modified (46)-(50) and (52)) is solved to transfer a feasibility cut to MP. After evaluating all NSs, if there is no unbounded DSP (i.e., 'a' is still equal to 0), for the feasible solution, the UB is calculated, and the NDSP (i.e., modified (45), (46)-(50) and (55)), is solved to obtain the POC. In MP, based on the obtained cuts, the investment cost of new constructed or bundled lines, on/off states of thermal units, and the charging/discharging states of ES, are optimized. After solving MP and calculating the LB, if the criterion of (53) is satisfied, the algorithm is terminated; otherwise, the next iteration is started. As shown in Fig. 1, the process in the red dashed box is executed for each set of decision variables when multiple feasible solutions are obtained for MP. Therefore, multiple cuts are transferred to MP.

- *The Proposed Contingency Screening Algorithm*

To identify the high-risk line outages for analyzing the *N-1* security criterion, a CS algorithm is developed, as shown in Fig. 2. In this algorithm, firstly, the outage of all existing, bundled, and new constructed lines is evaluated by solving the defined CSDSP. Note that the outage of each existing line is considered if the line is not bundled and there is no new constructed parallel line. Then, the loading index (LI) for each outage $l'$ is computed as equations (56) and (57). In equations (56) and (57) the LI is calculated under the outage of each existing and new constructed lines outage, respectively. The impact of bundling is modeled using $\psi_l$. The conditional operator of '|' is used to count the number of new and bundled lines just in the construction stages.

$$LI_{l'} = \frac{1}{L-1} \sum_{\substack{l \in \Omega_{el} \\ l \neq l'}} \left[ \frac{\max_{t \& h}\{|\overline{Pe_{t,l,h}}|\}}{\psi_l \cdot P_l^{max}} \right]^2 + \tag{56}$$

$$\frac{1}{\sum_{\substack{t \in \Omega_T, l \in \Omega_{nl}, c \in \Omega_C \\ |\overline{Y_{t,l,c}}| > \overline{Y_{t-1,l,c}}}} \overline{Y_{t,l,c}}} \sum_{l \in \Omega_{nl}, c \in \Omega_C} \left[ \frac{\max_{t \& h}\{|\overline{Pl_{t,l,c,h}}|\}}{P_l^{max}} \right]^2 \quad \forall l' \in \Omega_{el}$$

$$LI_{l'} = \frac{1}{L} \sum_{l \in \Omega_{el}} \left[ \frac{\max_{t \& h}\{|\overline{Pe_{t,l,h}}|\}}{\psi_l \cdot P_l^{max}} \right]^2 + \tag{57}$$

$$\frac{1}{\left(\sum_{\substack{t \in \Omega_T, l \in \Omega_{nl}, c \in \Omega_C \\ |\overline{Y_{t,l,c}}| > \overline{Y_{t-1,l,c}}}} \overline{Y_{t,l,c}}\right)-1} \sum_{\substack{l \in \Omega_{nl}, c \in \Omega_C \\ l \neq l'}} \left[ \frac{\max_{t \& h}\{|\overline{Pl_{t,l,c,h}}|\}}{P_l^{max}} \right]^2$$

$$\forall l' \in \Omega_{nl}$$

$$\boldsymbol{\psi_l} = 1 + \sum_{\substack{t \in \Omega_T, \ell \in \Omega_{bl} \\ |\overline{Yb_{t,\ell}}| > \overline{Yb_{t-1,\ell}}}} \left[ \left(Ab_\ell^l \cdot \overline{Yb_{t,\ell}}\right) \times \sum_{n \in \Omega_N} Ac_\ell^n \right]$$

After computing (56) or (57), the system load shedding due to each line outage (i.e., LSI) is calculated and normalized. Accordingly, the defined CS index in (58) is defined. It should be noted that for evaluating the severity of transmission line outages, the most important factor is the amount of relevant load shedding. Therefore, in CS index calculation based on (58), the risk of each outage $l'$ is measured as 20% of relevant LI plus 80% of relevant normalized load shedding.

$$CS_{l'} = 20\% \times (LI_{l'}) + 80\% \times \left(\widetilde{LSI_{l'}}\right) \tag{58}$$
$\sim$ : *Means normalized values*

$$LSI_{l'} = \sum_{t,i,h} LS_{t,i,h} \qquad \forall l' \in \Omega_{el} \text{ or } \forall l' \in \Omega_{nl}$$

Finally, the contingencies with CS index more than a threshold, e.g., $20\% \times \left[\max_{l'}(CS_{l'})\right]$, are selected as the high-risk NSs.

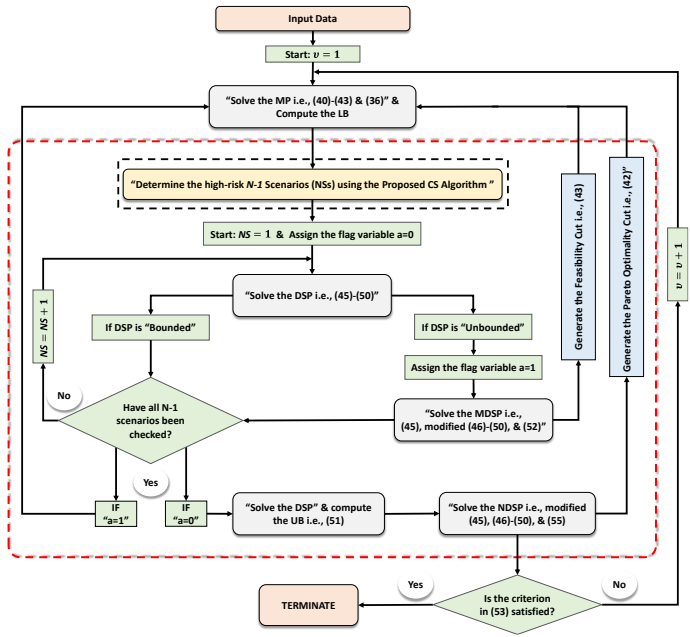

Fig. 1. The Overall Structure of the Proposed Model

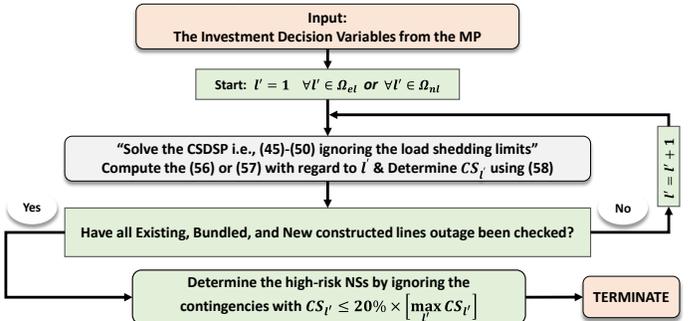

Fig. 2. The Structure of the Proposed Contingency Screening Algorithm

## IV. *Representative Hours*

To model the operational details in a long-term expansion planning model considering all 8760-time steps for each year, will result in a large unsolvable model. With this in mind, and to capture the uncertainties of load demand and WPP output power in an expansion planning model, time series aggregation methods can be utilized. Representative time periods can be extracted by clustering methods that are the most common methods for time series aggregation. In this paper, to decrease the computational complexity and capture the operational uncertainties of load and WPP output power, an accurate CTPC algorithm [12] is utilized. Using this algorithm, the proper representative hours for load demand and wind power historical data of Norway in 2019 [33] are extracted. Also, all data can be found in [34]. During the planning horizon, the CTPC algorithm is able to keep the chronology of parameters that are time dependent. Therefore, under the integration of ES devices and WPPs, the operational uncertainties can be captured more accurately in a long-term horizon. In Fig. 3 the utilized CTPC algorithm including eight main steps is described. Based on this algorithm, the chronology of data is captured by merging the adjacent clusters. By using the CTPC algorithm and considering the metrics presented in [35] to find the suitable number of representatives, the hourly load and wind power historical data are represented by 96 hours, as shown in Fig. 4. The extracted representative hours are illustrated across the year using each representative weight. Accordingly, instead of considering all hours of a year, the system uncertainties can be captured using the obtained 96 hours with less complexity. The real data, $Lf$, $Wf$, $\rho$, and aggregated representative hours are presented in [36].

## V. SIMULATION RESULTS

The proposed co-planning model is simulated over IEEE RTS 24-bus, and IEEE 118-bus test systems. The

---

**Start**

*Define the suitable number of representative hours, i.e., SH.*

**1**. Consider the data of load demand and WPP output power at each hour as an initial cluster, i.e., $h = 8760$, $h$ is the number of extracted clusters.

**2**. Calculate the centroid ($\bar{X}$) of each cluster $h$ as $\bar{X}_h = \frac{1}{|h|}\sum_{i \in h} x_i$, $|h|$ is the number of total elements in cluster $h$ and $x_i$ is element $i$ in cluster $h$.

**3**. Measure dissimilarity of each two adjacent clusters $h$ and $h'$ based on Euclidean distance of related centroids as $D(h, h') = \frac{2|h||h'|}{|h|+|h'|}\|\bar{X}_h - \bar{X}_{h'}\|^2$, in which $\|\bar{X}_h - \bar{X}_{h'}\|^2 = \sqrt{\sum_{i=1}^{N}(\bar{X}_{h_i} - \bar{X}_{h'_i})^2}$, $N$ is related centroids length.

**4**. Merge two clusters $\hat{h}$ and $\hat{h}'$ with the minimum dissimilarity.

**5**. Reduce the number of clusters by one, i.e., $h \to h - 1$.

**6**. If $h = SH$ go to step **7**, otherwise return to step **2**.

**7**. Calculate the final $h$ representatives using the related centroids $\bar{X}_h$.

**8**. Consider number of hours in each cluster as final weights $\rho_h$.

**End**

Fig. 3. Utilized chronological time-period clustering (CTPC) algorithm

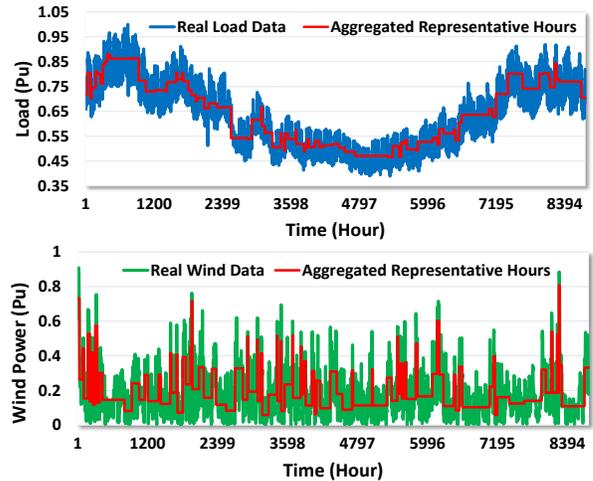

Fig. 4. Load and wind power real data and extracted representative hours

configuration and data of the test systems at the beginning of the planning horizon are considered as input data. For TEP problems the planning horizon is conventionally set from five to twelve years. Due to the complexity of TEP problems, the planning horizon can be divided into multi-year stages. It should be noted that without loss of generality, the planning horizon can be extended. In this paper, the planning time horizon is assumed to be 6 years, divided into 3 stages with a 2-years duration. The LT of all new lines, ES devices, and WPPs are considered 50, 10, and 20 years, respectively. In the case of single circuit lines, the investment cost of new lines, new two/four conductors per phase bundled lines, new RoW, and new substations are assumed as 1, 0.455/0.837, 0.034 $M\$/Km$, and 3.358 $M\$$, respectively. Moreover, in the case of double circuit lines, the above mentioned investment costs are 1.6, 2×(0.455/0.837), 1.142×(0.034) $M\$/Km$, and 2×(3.358) $M\$$, respectively [37]. The two and four conductors per phase bundled line can increase the rated power capacity up to 43% and 85%, respectively [22]. The cost of ES devices is considered as 500 $\$/Kw$ and 50 $\$/Kwh$ according to [14], and the charging and discharging efficiency are assumed to be 0.9. The degradation cost of ES is assumed as 5 $\$/Mwh$ [38]. The WPP investment cost is 2 $M\$/MW$, and the load shedding and wind curtailment cost are both considered 1000 $\$/Mwh$ [6]. The $\alpha$, $\beta$, and $\vartheta$ are assumed 15%, 50%, and $3h$, respectively, and when the *N-1* security criterion is ignored, both $\gamma$ and $\Phi$ are assumed to be zero. The yearly load growth and interest rate are both equal to 5%. All simulations are executed by CPLEX solver in GAMS using a PC with Intel Core i7, and 32 GB of RAM.

### A. *IEEE 24-bus Test System*

The modified IEEE 24-bus test system contains 30 existing single circuit and 4 double circuit lines. Three WPPs are candidates to be installed in buses 6, 14, and 20, each with a maximum capacity of 150 MW. Two WPPs are also assumed to be installed in new buses 25 and 26, both with 380 MW capacity. The main reason for introducing two new buses is considering and investigating the influence of connecting remote WPPs to the system through new corridors. The new bus 25 can be connected to buses 18 and 21 through two double circuit lines. Also, bus 26 can be connected to buses 16 and 19 through two double circuit lines. The capacity of WPPs is considered based on RPS policy that at the end of planning horizon, at least 15% of the system peak load should be supplied by WPPs. Moreover, the candidate existing buses for WPPs installation are the buses with a relatively high peak load demand. Five buses, i.e., 1, 6, 10, 25, and 26, are candidates for installing ES devices. The maximum power and energy capacity of each ES device are assumed as 200 MW and 1000 MWh, respectively [9]. The candidate buses for ES devices are also considered buses with a relatively high peak load demand and as near as possible to WPPs locations. In addition, fifteen new candidate lines along with four existing lines as the bundling candidates, are considered. It is possible to bundle each candidate existing line using two or four conductors per phase. It should be noted that to extract the proper new candidate lines, a static (one-stage) OPF is conducted over each case study in which the load demand in the last stage of planning horizon is considered. In addition, the possibility of load shedding in buses is also allowed in this static OPF. After executing the OPF, the candidate lines in existing corridors are the lines with a maximum loading of more than 65%, and all connected lines to the buses with a load shedding greater than zero. The utilized procedure considers the most congested paths as candidates, and the final decision is made based on economic and technical issues. All parameters and data of existing and candidate lines and thermal units can be found in [34].

The effectiveness of proposed accelerated BDD algorithm is evaluated over IEEE 24-bus test system using two designed schemes. In this regard, firstly, a normal BDD algorithm is used to solve the proposed model ignoring N-1 security criterion and load shedding possibility. In this normal BDD scheme, the proposed co-planning is conducted using new transmission lines and WPPs while the ES devices and bundling options are ignored. The result of this scheme is presented in Table II, as scheme '$N$'. Secondly, the proposed accelerated BDD algorithm is used to solve the considered

TABLE II
COMPARISON BETWEEN NORMAL BDD AND THE PROPOSED ACCELERATED BDD ALGORITHM

| BDD[1] | TIC (M$) | | | TOC (M$) |
|---|---|---|---|---|
| | NCL[2] | WPP | Total | 3978.218 |
| | | | | TPC (M$): |
| N[3] | 403.23 | 564.17 | 967.4 | Z=4945.618 |

**Line**: (16-19) **t=1**, (18-25)*, (16-26)* **t=2**, (6-10), (7-8), (15-21)* **t=3**
**WPP (MW): Bus 6**: 150, **Bus 14**: 136.268, **Bus 20**: 21.33 **t=1**,
**Bus 14**: 150, **Bus 20**: 150, **Bus 25**: 42.56, **Bus 26**: 185.693 **t=2**,
**Bus 25**: 317.784, **Bus 26**: 353.876 **t=3**
**Total Wind Curtailment**: 22296.037 MWh
**CPU time**≅ 6366 Sec

| A[4] | 403.23 | 564.17 | 967.4 | Z=4945.618 |

**Line**: (16-19) **t=1**, (18-25)*, (16-26)* **t=2**, (6-10), (7-8), (15-21)* **t=3**
**WPP (MW): Bus 6**: 150, **Bus 14**: 136.268, **Bus 20**: 21.33 **t=1**,
**Bus 14**: 150, **Bus 20**: 150, **Bus 25**: 42.56, **Bus 26**: 185.693 **t=2**,
**Bus 25**: 317.784, **Bus 26**: 353.876 **t=3**
**Total Wind Curtailment**: 22296.037 MWh
**CPU time**≅ 4887 Sec

**1**: Benders Dual Decomposition (BDD) Algorithm. **2**: New Constructed Line. **3**: Normal BDD Algorithm. **4**: Proposed Accelerated BDD Algorithm. *****: Double Circuit Line.

TABLE III
INTRODUCTION OF ALL DEFINED SCHEMES FOR IEEE 24-BUS TEST SYSTEM

| Schemes | New Lines | WPPs | BO[1] | ES | N-1 SC[2] | CS[3] | LSP[4] |
|---|---|---|---|---|---|---|---|
| I | ✓ | ✓ | ---- | ---- | ---- | ---- | ---- |
| II | ✓ | ✓ | ✓ | ---- | ---- | ---- | ---- |
| III | ✓ | ✓ | ✓ | ✓ | ---- | ---- | ---- |
| IV | ✓ | ✓ | ✓ | ✓ | ✓ | ---- | ---- |
| V | ✓ | ✓ | ✓ | ✓ | ✓ | ✓ | ---- |
| VI | ✓ | ✓ | ✓ | ✓ | ✓ | ✓ | ✓ |

**1**: Bundling Option. **2**: Security Criterion. **3**: Contingency Screening.
**4**: Load Shedding Possibility.

model in scheme '$N$'. The result of this scheme is also presented in Table II, as scheme '$A$'. The obtained result confirms the effectiveness of proposed accelerated BDD algorithm in solving the proposed co-planning model. The computation time is reduced more than 23% by using the proposed accelerated BDD algorithm. Therefore, the accelerated BDD algorithm is utilized for other simulation schemes. Six different schemes are defined to show the effectiveness of the proposed model over IEEE 24-bus test system, as defined in Table III. In schemes I, II & III the $N-1$ security criterion and load shedding possibility are ignored. In scheme I, the proposed co-planning is conducted using new transmission lines and WPPs while the ES devices and bundling options are ignored. Scheme II is similar to scheme I with incorporating the bundling option. In scheme III, all planning options (i.e., new lines, WPPs, ES devices, and bundling) are considered. In schemes IV, V, & VI the results of the proposed co-planning model considering the $N-1$ security criterion are discussed. In scheme IV, all $N-1$ scenarios are evaluated, ignoring the load shedding possibility. In scheme V, according to the proposed CS algorithm, the $N-1$ scenarios are limited to the obtained high-risk ones ignoring the load shedding possibility. Scheme VI is similar to scheme V except that the possibility of load shedding is considered with $\gamma$ and $\Phi$ equal to 20% and 15%, respectively. In Table III, all schemes are introduced in order to provide more clarification.

The results of defined schemes I, II, and III are presented in Table IV. In Table V, the results of defined schemes IV, V, and VI are reported. Moreover, in Fig. 5, the results of scheme III are illustrated. According to Table IV, when the $N-1$ security criterion is ignored (i.e., schemes I, II, & III), the obtained results for scheme III present an optimal co-planning configuration with a more economical total planning cost (TPC). In scheme III, two double-circuit lines are constructed between buses 18-25 and 16-26 at the second stage, and a double-circuit line is constructed in corridor 15-21 at the last stage of the planning horizon. In scheme III, the TIC, TOC, and TPC are 1160.212, 3644.315, and 4804.527 $M\$$, respectively. Therefore, considering the bundling and ES devices options in scheme III results in 141.091 $M\$$ cost saving compared to scheme I. Unlike scheme I, in scheme III, instead of constructing two new single-circuit lines in corridors 16-19, and 7-8, the existing lines are bundled with less expensive cost, which confirms the bundling option effectiveness. The total wind curtailment of 22296.037 MWh in schemes I and II is reduced to zero in schemes III by

TABLE IV
RESULTS OF SCHEMES I, II AND III OF THE PROPOSED CO-PLANNING MODEL OVER IEEE 24-BUS TEST SYSTEM

| Scheme | NCL[1] | TIC (M$) WPP | ES | BL[2] | Total | TOC (M$) 3978.218 TPC (M$): |
|---|---|---|---|---|---|---|
| I | 403.23 | 564.17 | ---- | ---- | 967.4 | Z=4945.618 |

**Line**: (16-19) **t=1**, (18-25)*, (16-26)* **t=2**, (6-10), (7-8), (15-21)* **t=3**
**WPP (MW): Bus 6**: 150, **Bus 14**: 136.268, **Bus 20**: 21.33 **t=1**,
**Bus 14**: 150, **Bus 20**: 150, **Bus 25**: 42.56, **Bus 26**: 185.693 **t=2**,
**Bus 25**: 317.784, **Bus 26**: 353.876 **t=3**
**Total Wind Curtailment**: 22296.037 MWh

| | | | | | | 3952.159 |
|---|---|---|---|---|---|---|
| II | 368.422 | 564.17 | ---- | 22.827 | 955.419 | Z=4907.578 |

**Line**: (18-25)*, (16-26)* **t=2**, (6-10), (15-21)* **t=3**
**WPP (MW): Bus 6**: 150, **Bus 14**: 7.6, **Bus 20**: 150 **t=1**,
**Bus 14**: 150, **Bus 25**: 42.5, **Bus 26**: 185.7 **t=2**,
**Bus 25**: 317.78, **Bus 26**: 353.87 **t=3**
**Bundling with 2-Conductors per Phase**: (16-19) **t=1**, (7-8) **t=2**
**Total Wind Curtailment**: 22296.037 MWh

| | | | | | | 3644.315 |
|---|---|---|---|---|---|---|
| III | 353.167 | 564.17 | 227.558 | 15.317 | 1160.212 | Z=4804.527 |

**Line**: (18-25)*, (16-26)* **t=2**, (15-21)* **t=3**
**WPP (MW): Bus 6**: 150, **Bus 14**: 7.6, **Bus 20**: 150 **t=1**,
**Bus 14**: 150, **Bus 25**: 25.101, **Bus 26**: 203.152 **t=2**,
**Bus 25**: 330.256, **Bus 26**: 341.404 **t=3**
**ES (MW, MWh): Bus 6**: 175.344, 526.032, **Bus 10**: 200, 600 **t=1**,
**Bus 25**: 114.95, 344.85, **Bus 26**: 113.2, 339.6 **t=2**
**Bundling with 2-Conductors per Phase**: (16-19) **t=1**, (7-8) **t=3**
**Total Wind Curtailment**: 0

**1:** New Constructed Line. **2:** Bundled Line. *****: Double Circuit Line.

utilizing ES devices. Also, in scheme I the constructed line between buses 6-10 is ignored in scheme III due to installing ES devices in bus 6. The presented comparisons confirm the beneficial impacts of ES devices and transmission line bundling on the TEP problem under the high penetration of WPPs. Compared to scheme I, in scheme II by considering the bundling option in the proposed model, instead of installing a new transmission line between buses 16-19, this line is bundled by using two conductors per phase in the first stage. Moreover, in scheme II the transmission line between buses 7-8 is bundled in stage two using two conductors per phase. Therefore, in comparison to scheme I, the construction of a new line in corridor 7-8 is avoided in scheme II. The TPC in scheme II is 4907.578 *M$* which results in 38.04 *M$* cost saving compared to scheme I. The comparison between the results of schemes I and II confirms the impact of bundling option on TPC reduction. In scheme III, TPC is 103.051 *M$* less expensive than scheme II which confirms the influence of ES devices on TPC reduction. The installation stage and capacities of WPPs and ES devices are presented in a cumulative format in Tables IV and V. Fig. 6 illustrates the shares of output energy of the thermal units, WPPs, and ES devices in the total energy supply at each stage for scheme III. As shown, the WPPs penetration is increased in each stage regarding the considered RPS policy. This increment led to a decrease in thermal units share that declined the TOP. The share of ES is defined as the difference between total discharging and charging energy.

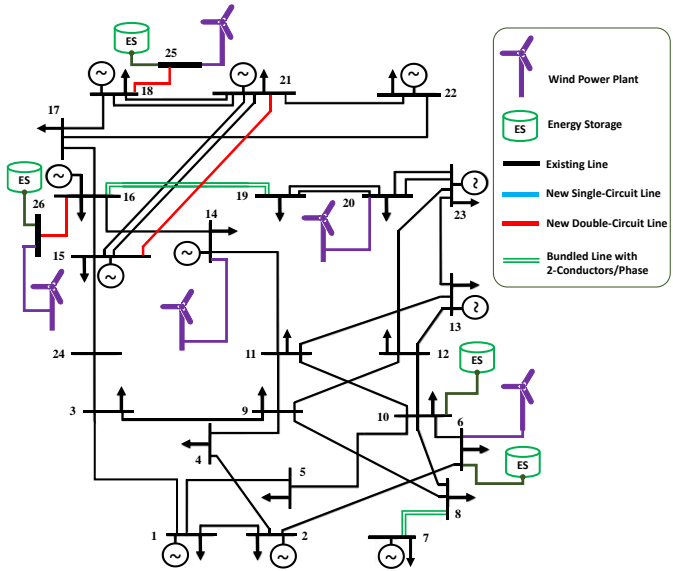

Fig. 5. Results of the scheme III for IEEE 24-bus test system

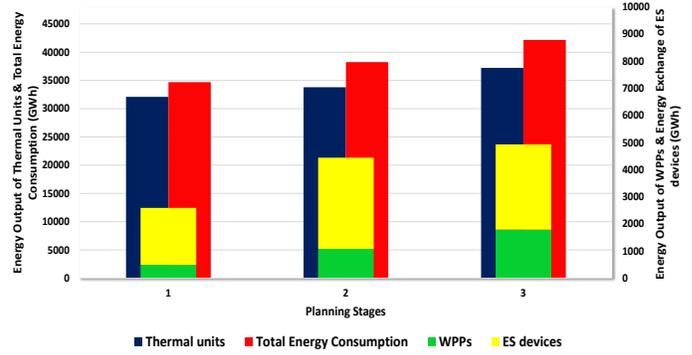

Fig. 6. The total energy supply at each stage of scheme III for IEEE 24-bus test system

By modeling the *N-1* security criterion (i.e., schemes IV, V, & VI), according to the presented results of scheme IV in Table V, the TIC, TOC, and TPC are 1332.828, 3586.366, and 4919.194 *M$*, respectively. In this scheme, three double-circuit and five single-circuit lines, along with four installed ES devices in buses 6, 10, 25, and 26 are needed to ensure the *N-1* security criterion without any wind curtailment and load shedding. In comparison to scheme III, five more single-circuit lines in corridors 7-8, 6-10, 16-19, 3-24, and 15-24 are constructed in scheme IV. Although with respect to scheme III, the TOC of scheme IV is reduced up to 57.949 *M$*, the TIC and TPC are respectively 172.616 and 114.667 *M$* more expensive than scheme III. It shows the significant impact of considering the *N-1* security criterion. In scheme V, by considering just the extracted high-risk NSs using the CSDSP, the obtained results show about 47.77% saving in computation time without any deviation from the results of scheme IV, which confirms the effectiveness of the proposed CS algorithm. In scheme VI, the TPC is 90.033 *M$* less expensive than schemes IV and V. In this scheme, the total load shedding during the planning horizon is 1748.966 MWh, which is due to an expected load shedding up to 1126.478 MWh in bus 16, and 622.488 MWh in bus 19. In scheme VI, two double-circuit and two single-circuit lines are constructed without any wind curtailment. The existing line between buses 16-19 is bundled

TABLE V
RESULTS OF SCHEMES IV, V, AND VI OF THE PROPOSED CO-PLANNING MODEL OVER IEEE 24-BUS TEST SYSTEM

| Scheme | TIC (M$) | | | | | TOC (M$) |
| --- | --- | --- | --- | --- | --- | --- |
| | NCL | WPP | ES | BL | Total | 3586.366 |
| | | | | | | TPC (M$): |
| IV | 530.31 | 564.17 | 238.349 | ---- | 1332.828 | **Z=4919.194** |

*Line*: (7-8), (6-10), (16-19) **t=1**, (18-25)*, (16-26)*, (15-21)* **t=2**, (3-24), (15-24) **t=3**
*WPP* (MW): **Bus 6**: 150, **Bus 14**: 7.6, **Bus 20**: 150 **t=1**,
**Bus 14**: 36.176, **Bus 25**: 145.953, **Bus 26**: 196.124 **t=2**,
**Bus 14**: 150, **Bus 25**: 330.256, **Bus 26**: 341.404 **t=3**
*ES* (MW, MWh): **Bus 6**: 200, 600, **Bus 10**: 200, 600 **t=1**,
**Bus 25**: 114.95, 344.85, **Bus 26**: 113.2, 339.6 **t=2**
*Total Wind Curtailment*: 0
*CPU time* ≅ 27800 Sec.

| V | 530.310 | 564.17 | 238.349 | ---- | 1332.828 | 3586.366 |
| --- | --- | --- | --- | --- | --- | --- |
| | | | | | | **Z=4919.194** |

*Line*: (7-8), (6-10), (16-19) **t=1**, (18-25)*, (16-26)*, (15-21)* **t=2**, (3-24), (15-24) **t=3**
*WPP* (MW): **Bus 6**: 150, **Bus 14**: 7.6, **Bus 20**: 150 **t=1**,
**Bus 14**: 36.176, **Bus 25**: 145.953, **Bus 26**: 196.124 **t=2**,
**Bus 14**: 150, **Bus 25**: 330.256, **Bus 26**: 341.404 **t=3**
*ES* (MW, MWh): **Bus 6**: 200, 600, **Bus 10**: 200, 600 **t=1**,
**Bus 25**: 114.95, 344.85, **Bus 26**: 113.2, 339.6 **t=2**
*Total Wind Curtailment*: 0
*CPU time* ≅ 14520 Sec.

| VI | 390.026 | 564.17 | 238.349 | 21.036 | 1213.581 | 3615.58 |
| --- | --- | --- | --- | --- | --- | --- |
| | | | | | | **Z=4829.161** |

*Line*: (6-10) **t=1**, (18-25)*, (16-26)* **t=2**, (16-19) **t=3**
*WPP* (MW): **Bus 6**: 150, **Bus 14**: 7.6, **Bus 20**: 150 **t=1**,
**Bus 14**: 150, **Bus 25**: 25.101, **Bus 26**: 203.152 **t=2**,
**Bus 25**: 330.256, **Bus 26**: 341.404 **t=3**
*ES* (MW, MWh): **Bus 6**: 200, 600, **Bus 10**: 200, 600 **t=1**,
**Bus 25**: 114.95, 344.85, **Bus 26**: 113.2, 339.6 **t=2**
*Bundling with 2-Conductors per Phase*: (16-19) **t=1**
*Bundling with 4-Conductors per Phase*: (7-8) **t=3**
*Total Wind Curtailment*: 0     &     *Total Load Shedding*: 1748.966 MWh
*CPU time* ≅ 7510 Sec.

by using two conductors per phase. Also, by using four conductors per phase, the existing line between buses 7-8 is bundled. The load shedding possibility decreases the TIC and TPC and increases the TOC.

### B. IEEE 118-bus Test System

The IEEE 118-bus test system includes 172 existing single and 7 double circuit transmission lines. Ten WPPs are candidates to be installed in existing buses 6, 12, 20, 34, 54, 70, 90, 112, 119, and 120, all with a maximum 180 MW capacity. Two WPPs with a maximum capacity of 400 MW are considered to be installed in new buses 119 and 120. The new buses 119, and 120 can be connected to the system through buses 59, and 116, respectively. Moreover, thirty new candidate lines along with seven candidate existing lines for bundling, are considered. Like IEEE 24-bus test system, it is possible to bundle each candidate existing line using two or four conductors per phase. Ten ES devices with a maximum power and energy capacity of 200 MW and 1000 MWh, are assumed as candidates for installation in buses 3, 12, 22, 32,

TABLE VI
INTRODUCTION OF ALL DEFINED SCHEMES FOR IEEE 118-BUS TEST SYSTEM

| Schemes | New Lines | WPPs | BO | ES | *N-1* SC | CS |
| --- | --- | --- | --- | --- | --- | --- |
| I | ✓ | ✓ | ---- | ---- | ---- | ---- |
| II | ✓ | ✓ | ✓ | ✓ | ---- | ---- |
| III | ✓ | ✓ | ✓ | ✓ | ✓ | ✓ |

55, 62, 77, 92, 119 and 120. All parameters and data of IEEE 118-bus test system are available in [34]. As presented in Table VI, for this test system three schemes are defined. Three schemes are not repeated for IEEE 118-bus test system with regard to the defined schemes for IEEE 24-bus test system. Indeed, schemes II, IV, and VI of Table III are not repeated in Table VI. Generally, the considered schemes for IEEE 118-bus in Table VI cover the other schemes. As presented in Table VI, in schemes I & II, the simulation results are provided without considering the *N-1* security criterion. In scheme I, the proposed co-planning model is examined ignoring the ES devices and bundling options. In scheme II, all planning options are considered. In schemes III the results of the proposed co-planning model considering the *N-1* security criterion and the proposed CS algorithm are discussed. In all schemes, the load shedding possibility is ignored. The numerical results for defined schemes are presented in Table VII. In scheme I, six single and two double circuit new lines are constructed and the TIC, TOC, and TPC are 1290.136, 8630.883, and 9921.019 M$. In this scheme, total wind energy of 5923.024 MWh is curtailed. It should be noted that the cost of new constructed transmission lines depends on the lines length. In scheme II, in which both ES devices and bundling options are incorporated, three single and two double circuit new lines are constructed. In comparison to scheme I, in scheme II instead of constructing new lines in corridors 77-78, and 37-39, the existing lines are bundled using four conductors per phase, in the first and last stages. In addition, the existing lines between buses 17-18, and 23-32 are bundled in the first and last stages of the planning horizon using two conductors per phase in scheme II. The TIC, TOC, and TPC are 1511.452, 8290.353, and 9801.805 M$ in scheme II, which shows a 119.214 M$ cost saving in TPC compared to scheme I. This total cost saving confirms the effectiveness of the proposed model in reducing the cost of TEP problem. In other words, the obtained numerical results show that the ES devices and bundling options have a noticeable influence on achieving an optimal expansion plan. In scheme II, the total wind curtailment is reduced to 343.332 MWh that is mainly due to considering ES devices. In scheme III, the *N-1* security criterion and the proposed CS algorithm are considered. In this scheme seventeen single and five double circuit lines are constructed to make the system *N-1* secure. The transmission lines between buses 23-32, and buses 22-23 are bundled in stage two by using respectively two and four conductors per phase. In scheme III, the TIC, TOC, and TPC are 2077.626, 8150.142, and 10227.768 M$. Although in comparison to scheme II, the TOC of scheme III is reduced up to 140.211 M$, the TIC and TPC are 566.174 and 425.963 M$ more expensive than scheme II. The more utilization of ES devices in scheme III compared to scheme II, results in a zero total

TABLE VII
RESULTS OF SCHEMES I, II, AND III OF THE PROPOSED CO-PLANNING MODEL OVER IEEE 118-BUS TEST SYSTEM

| Scheme | TIC (M$) | | | | | TOC (M$) |
|---|---|---|---|---|---|---|
| | NCL | WPP | ES | BL | Total | 8630.883 |
| I | 262.596 | 1027.54 | ---- | ---- | 1290.136 | TPC (M$): Z=9921.019 |

**Line**: (77-78) **t=1**, (69-77) **t=2**,
(5-11), (37-39), (49-51), (94-95), (59-119)*, (116-120)* **t=3**
**WPP (MW): Bus 6**: 84.85, **Bus 12**: 180, **Bus 20**: 180,
**Bus 54**: 25.5, **Bus 112**: 90 **t=1**,
**Bus 6**: 180, **Bus 34**: 180, **Bus 54**: 180, **Bus 90**: 155.35, **Bus 112**: 180 **t=2**,
**Bus 70**: 180, **Bus 90**: 180, **Bus 119**: 400, **Bus 120**: 202.92 **t=3**
**Total Wind Curtailment**: 5923.024 MWh

| II | 227.198 | 1027.54 | 225.192 | 31.522 | 1511.452 | 8290.353 Z=9801.805 |
|---|---|---|---|---|---|---|

**Line**: (69-77) **t=2**, (5-11), (30-38), (59-119)*, (116-120)* **t=3**
**WPP (MW): Bus 6**: 180, **Bus 12**: 180, **Bus 20**: 61.53,
**Bus 54**: 49.54, **Bus 112**: 90 **t=1**,
**Bus 20**: 180, **Bus 34**: 41, **Bus 54**: 180, **Bus 70**: 180, **Bus 90**: 157.93,
**Bus 112**: 136.43 **t=2**,
**Bus 34**: 180, **Bus 90**: 180, **Bus 112**: 180, **Bus 119**: 367.6, **Bus 120**: 235.3 **t=3**
**ES (MW, MWh): Bus 12**: 97.12, 291.36, **Bus 22**: 13.85, 41.55,
**Bus 32**: 76.225, 228.675, **Bus 55**: 52.715, 158.145, **Bus 77**: 71.41, 214.23,
**Bus 92**: 125, 375 **t=1**,
**Bus 12**: 106.34, 319.02, **Bus 22**: 35.65, 106.96 **t=2**,
**Bus 12**: 138.58, 415.75, **Bus 119**: 97.05, 291.15, **Bus 120**: 64.95, 194.85 **t=3**
**Bundling with 2-Conductors per Phase**: (17-18) **t=1**, (23-32) **t=3**
**Bundling with 4-Conductors per Phase**: (77-78) **t=1**, (37-39) **t=3**
**Total Wind Curtailment**: 343.332 MWh

| III | 705.105 | 1027.54 | 296.156 | 48.825 | 2077.626 | 8150.142 Z=10227.768 |
|---|---|---|---|---|---|---|

**Line**: (5-8), (37-39), (42-49)*, (49-50), (49-51), (60-61), (63-64), (77-78), (89-90)*, (68-116), (12-117) **t=1**, (5-11), (59-63), (64-65), (69-77), (77-78) **t=2**,
(9-10), (30-38), (49-66)*, (79-80), (59-119)*, (116-120)* **t=3**
**WPP (MW): Bus 6**: 160.132, **Bus 12**: 180, **Bus 20**: 125.8, **Bus 112**: 94.3 **t=1**,
**Bus 20**: 180, **Bus 34**: 180, **Bus 54**: 180, **Bus 70**: 180, **Bus 90**: 80.9 **t=2**,
**Bus 6**: 180, **Bus 90**: 180, **Bus 112**: 180, **Bus 119**: 365.7, **Bus 120**: 237.217 **t=3**
**ES (MW, MWh): Bus 12**: 77.62, 232.86, **Bus 22**: 31.16, 93.5,
**Bus 32**: 117.92, 353.77, **Bus 55**: 74.73, 224.2, **Bus 77**: 76.96, 230.9,
**Bus 92**: 111.91, 335.73 **t=1**,
**Bus 12**: 167.66, 503, **Bus 22**: 78.53, 235.6, **Bus 55**: 126.2, 378.6,
**Bus 77**: 89.55, 268.66, **Bus 92**: 122.7, 368.1 **t=2**,
**Bus 119**: 105.77, 317.32, **Bus 120**: 66.18, 198.55 **t=3**
**Bundling with 2-Conductors per Phase**: (23-32) **t=2**
**Bundling with 4-Conductors per Phase**: (22-23) **t=2**
**Total Wind Curtailment**: 0
**CPU time** ≅ 63162 Sec.

wind curtailment.

## VI. CONCLUSION

This paper presented a secure expansion co-planning of transmission lines, ES devices, and WPPs considering bundling options. A CS algorithm was utilized to identify the high-risk *N-1* contingency scenarios. Also, the load demand and wind power uncertainties were captured using a CTPC approach. The major findings of this work are summarized as follows. 1) The integration of ES devices in the TEP model defers the transmission investment, relieves transmission congestion, and facilitates renewable integration with less wind curtailment. 2) The planning of WPP concerning the RPS policy goals declines the share of thermal units in supplying the load and the total operation cost. 3) The *N-1* security criterion impacts the co-planning expansion plan significantly. Considering the scheduled value of load shedding reduces the TPC. Such load shedding can be planned to be realized via demand response programs. The proposed CS algorithm reduces the secure co-planning model computational burden without deviating from the original secure plan. 4) Due to the high cost of new RoW and environmental restrictions, upgrading the existing lines using the bundling option results in an economic TEP and avoids new lines. 5) The proposed accelerated BDD algorithm can handle different parts of the proposed co-planning model with efficient consideration of single high-risk contingencies via the *N-1* criterion. This paper addressed the expansion planning of ES devices and WPPs in a centralized structure. In future works, this co-planning can be discussed in the electricity market environment. Moreover, modeling and evaluating ES device contingencies can be addressed in future works.